\documentclass[aps,pra,preprint,groupedaddress]{revtex4-1}
\usepackage{amsmath}

\ifx\pdfoutput\undefined
\usepackage[dvips]{graphicx}
\else
\usepackage[pdftex]{graphicx}
\fi
\usepackage{epstopdf}

\begin{document}
\newcommand{\be}{\begin{equation}}
\newcommand{\ee}{\end{equation}}

\title{Disorder effects in tunable waveguide arrays with parity-symmetric tunneling}
\author{Clinton Thompson}
\affiliation{Department of Physics, Indiana University Purdue University Indianapolis (IUPUI), Indianapolis, Indiana 46202, USA}
\author{Yogesh N. Joglekar}
\affiliation{Department of Physics, Indiana University Purdue University Indianapolis (IUPUI), Indianapolis, Indiana 46202, USA}
\author{Gautam Vemuri}
\affiliation{Department of Physics, Indiana University Purdue University Indianapolis (IUPUI), Indianapolis, Indiana 46202, USA}
\date{\today}

\begin{abstract}
We investigate the effects of disorder on single particle time-evolution and two-particle correlations in an array of evanescently coupled waveguides with position-dependent tunneling rates. In the clean limit, the energy spectrum of such an array is widely tunable. In the presence of a Hermitian on-site or tunneling disorder, we find that the localization of a wave packet is highly sensitive to this energy spectrum. In particular, for an input confined to a single waveguide, we show that the fraction of light localized to the original waveguide depends on the tunneling profile. We compare the two-particle intensity correlations in the presence of Hermitian, tunneling disorder and non-Hermitian, parity-and-time-reversal ($\mathcal{PT}$) symmetric, on-site potential disorder. We show the two-particle correlation function in both cases is qualitatively similar, since both disorders preserve the particle-hole symmetric nature of the energy spectrum. 
\end{abstract}
\maketitle

\section{Introduction}
\label{sec:intro}

The development of integrated photonic structures in recent years has allowed researchers to study many condensed matter effects in optical systems. For example, the motion of an electron in a periodic potential has been translated into the propagation of light in a waveguide array and Bloch-oscillation-like phenomena have been directly observed \cite{Peschel1998,Rai2009,Bromberg2010}. Optical waveguide arrays have also attracted considerable attention because the diffraction properties of light in these structures are very different from that in bulk media \cite{Christodoulides2003}.  

In recent years, arrays of evanescently coupled waveguides have attracted considerable interest as versatile structures suitable for the study of quantum and condensed matter phenomena such as Bloch oscillations, Zener tunneling \cite{Trompeter2006}, Dirac zitterbewegung \cite{Longhi2010a}, quantum random walks \cite{Perets2008}, and quantum ratchets \cite{Thompson2011}. This interest is a result of being able to observe quantum phenomena on a length-scale of a few millimeters, the typical length of a waveguide, as the discrete Schr\"{o}dinger equation for a particle on a lattice is identical to the longitudinal component of the Helmholtz equation for the electric field in an array of waveguides \cite{Lahini2008,Jones1965}. Non-classical states of light have also been used to study quantum properties of light propagating through waveguide arrays \cite{Rai2009,Thompson2010,Thompson2011,Bromberg2010}. In particular, the quantum correlations of two non-interacting indistinguishable  particles propagating simultaneously have shown the effect of their initial separation on propagation of the particles in clean \cite{Lahini2010} and disordered \cite{Peruzzo2010} systems.
 
Over the last decade, since the seminal work of Bender and co-workers \cite{Bender1998, Bender2002}, there has been much interest in the study of systems that are described by non-Hermitian, but parity- $(\mathcal{P})$ and time-reversal- $(\mathcal{T})$ symmetric Hamiltonians. This symmetry has led to predictions of new phenomenon such as Bloch oscillations in complex crystals \cite{Longhi2009}, an optical medium that can simultaneously act as an emitter and a perfect absorber of coherent waves \cite{Longhi2010c}, and induced quantum coherence between Bose-Einstein condensates \cite{Xiong2010}. $\mathcal{PT}$-symmetry breaking in a classical system has recently been experimentally observed in waveguide arrays \cite{Guo2009,Ruter2010}. Some of the predictions for waveguide arrays with $\mathcal{PT}$-symmetric Hamiltonians are Bloch oscillations in arrays with defects \cite{Longhi2010d}, the invisibility of defects in such an array \cite{Longhi2010b}, directed transport in non-linear arrays \cite{Ramezani2010}, periodic wave packet reconstruction \cite{Longhi2010}, and double refraction \cite{Makris2008}. Most of the waveguide arrays considered in these works are uniform, meaning the evanescent coupling between adjacent waveguides is the same. However, over the past year, a strong interest in waveguide arrays with position-dependent tunneling rate or evanescent coupling has developed~\cite{Longhi2010,Joglekar2011,Joglekar2011a}; this interest is driven, in part, by the tremendous flexibility that is offered by optical waveguides where the evanescent coupling can be varied by two orders of magnitude, and by the tunable properties of wave packet evolution in such arrays with non-uniform, position-dependent tunneling rates~\cite{Joglekar2011a}. 

Recent theoretical and experimental reports have explored the role of disorder in coupled waveguide arrays, and demonstrated Anderson localization of light in one-dimension. However, there are still several fundamental questions that remain unresolved with regard to the localization phenomenon (see \cite{Jovic2011}). For example, how is the localization defined in a lattice which departs from the standard configuration in which the tunneling amplitudes are equal?  By studying different models of non-uniform tunneling, one can probe the nature of localization.

Finally, note that although there is an exact mapping between coupled optical waveguides and a one-dimensional lattice model, qualitatively, coupled optical waveguides allow exploration of dynamics and disorder effects in a parameter regime that is virtually inaccessible in their condensed matter counterparts: the number of waveguides $N \sim 10-100$  is much smaller than typical number of lattice sites $N\gtrsim 10^{6}$ in any quantum wire; the typical disorder strength $\Delta$ in optical waveguides is  comparable to the bandwidth, whereas in condensed matter systems, $\Delta\lesssim E_{F} \ll$ bandwidth; the typical wavepacket in coupled waveguides spans the entire bandwidth whereas in condensed matter systems, a typical wavepacket only spans energies comparable to $E_{F}\ll $ bandwidth. Thus, it is crucially important to quantitatively define and explore the interaction between dynamics and disorder in lattice models with a position-dependent tunneling amplitude that are applicable to optical waveguide lattices. 

The inclusion of position-dependent tunneling rates, and the presence of disorder, makes our problem analytically intractable.  Therefore, most studies on disorder effects on light propagation in waveguide arrays are numerical in nature.  The few exceptions are the works on Bloch oscillations and related effects, in which the disorder is absent, and the tunneling amplitudes are constant.  Since the fabrication of waveguides with non-uniform tunneling rates is technologically feasible, as is the inclusion of disorder in such arrays, it is important to investigate the evolution of light in such systems.  As we show in this paper, the results are dramatically different from what one finds in systems with constant tunneling.

In this paper, we investigate the effect of disorder in waveguide arrays with a position-dependent tunneling rate $C(j)$. We study the time evolution of a wave packet and the effect the tunneling rate's functional form has upon the disorder-induced localization of the wave packet.  The appropriate functional form of the tunneling function increases the localization due to disorder, but, for a large disorder strength, the fraction of light localized to the initial waveguide may be independent of the global structure of the array. We find that, based on the initial wave function profile, disorder can cause broadening or localize the wave packet to two different waveguides. We compare and contrast the intensity-intensity correlations for random, Hermitian, off-diagonal disorder and random, non-Hermitian, $\mathcal{PT}$-symmetric, loss and gain disorder. We show that although both disorders give rise to qualitatively identical correlation functions, the full correlation matrix can distinguish between the two. 

The plan of the paper is as follows. In Sec.~\ref{sec:model} we introduce the tight-binding lattice Hamiltonian, the tunneling functions, and the disorder we consider. Sec.~\ref{sec:dis} contains results for the competition between tunneling and disorder, and how it affects the time-evolution and localization of a wave packet. In Sec.~\ref{sec:hbt} we extend the model to include non-Hermitian, $\mathcal{PT}$-symmetric, loss and gain disorder, and present the results for the intensity-intensity correlations that arise from two distinct disorders. We conclude the paper with a brief discussion in Sec.~\ref{sec:disc}.


\section{Tight-binding Model}
\label{sec:model}
We consider an array of $N$ single-mode, evanescently coupled waveguides where the propagation of light is described by a tight-binding Hamiltonian. In second quantized form, the Hamiltonian is given by 
\be
\label{eq:Hamiltonian} 
\hat{H} = \hbar\sum_{j=1}^{N}\beta_{j}a_{j}^{\dagger}a_{j}+\hbar\sum_{j=1}^{N-1}C(j)(a_{j+1}^{\dagger}a_{j}+a_{j}^{\dagger}a_{j+1})
\ee  
where $a_{j}^{\dagger}$ ($a_j$) represents the creation (annihilation) operator for a photon in waveguide $j$, $\beta_{j}$ is the linear-propagation constant (or equivalently the potential) at site $j$, and $C(j)$ is the tunneling rate between waveguides $j$ and $j+1$. A uniform array is characterized by a constant tunneling rate, $C(j)= C_{0}$. We choose a parity-symmetric tunneling function characterized by a single parameter $\alpha$~\cite{Joglekar2011a}
\be 
\label{eqn:tunnelingrate} 
C_\alpha(j)=C_{0}[j(N-j)]^{\alpha/2}.
\ee
Experimentally, one can engineer this tunneling function by symmetrically increasing (decreasing) the center-to-center distance between adjacent waveguides for negative (positive) values of $\alpha$. When $\alpha>0$, the tunneling rate is maximum at the center of the array whereas for $\alpha<0$, it is minimum. The energy-spectrum bandwidth $\Delta_{\alpha}$ is defined as $\Delta_{\alpha}=E_{\max}-E_{\min}$ where $E_j$ are eigenvalues of the Hamiltonian~(\ref{eq:Hamiltonian}); for a clean system, it scales as $\Delta_{\alpha}(N)\sim\hbar C_{0}N^{\alpha}$ for $\alpha \geq 0$ and $\Delta_{\alpha} \sim \hbar C_{0}N^{-|\alpha|/2}$ for $\alpha < 0$~\cite{Joglekar2011}. We choose the bandwidth inverse as the characteristic time-scale for the system, $\tau_{\alpha}(N)=\hbar/\Delta_{\alpha}(N)$. 

We introduce a Hermitian disorder through random variations of the on-site potential $\beta_j$. The mean value of on-site potential is irrelevant as long as it is the same for each waveguide; it only introduces a constant shift of the energy spectrum, and does not affect the bandwidth or the characteristic time. Therefore, it is set to zero. Since the disorder-induced localization is independent of the probability distribution of disorder provided the different distributions have zero mean and the same variance \cite{Thompson2010}, we use a Gaussian distribution, 
\be 
\label{eqn:distribution}
P(\beta_j)=\frac{1}{\sqrt{2\pi\sigma^{2}}}\exp\left(\frac{-\beta_j^{2}}{2\sigma^{2}}\right)
\ee
where $\sigma^{2}$ is the variance of the distribution, and thus, $\hbar\sigma$ characterizes the strength of the disorder. Note that, in contrast with the waveguide-dependent tunneling rate, the variance of disorder is {\it not dependent} upon the waveguide index $j$. 

Starting with an arbitrary initial state $|\psi(0)\rangle$ and the Hermitian Hamiltonian $\hat{H}_\alpha$, Eq.(\ref{eq:Hamiltonian}), with given on-site disorder potentials, we obtain the unitary time-evolution operator $\hat{U}(t)=\exp(-i\hat{H}t/\hbar)$, the time-evolved wave function $|\psi(t)\rangle=\hat{U}(t)|\psi(0)\rangle$, and the time-evolved, waveguide-index dependent intensity $I(j,t)=|\langle j|\psi(t)\rangle|^2$. We consider an array with $N\gg 1$ waveguides, and use Box-Muller algorithm to generate the random disorder with zero mean and the desired variance \cite{Box1958}. The characteristic time $\tau_\alpha$ and the energy bandwidth $\Delta_{\alpha}$ change with each disorder realization, and disorder averaging is carried out over $N_r\sim 10^3-10^6$ realizations to ensure that the results are independent of the number of realizations.  

\section{Wavepacket evolution in the presence of disorder}  
\label{sec:dis}

In a uniform waveguide array, a weak disorder exponentially localizes a wave packet that is initially confined to a single waveguide~\cite{Lahini2008}.  We begin by investigating the effect of position-dependent tunneling function $C_\alpha(j)$ on the localization of light that is initially confined to a single waveguide by calculating the $\alpha$-dependence of the fraction of total intensity that remains in the initial waveguide at times $t\gg\tau_\alpha$ for varying strengths of disorder. Similar to the energy-scale, disorder $\hbar\sigma$ is measured in units of clean-system bandwidth $\Delta_{\alpha}^{(0)}$, although, we will see below that the question of whether to use a global-scale, such as the bandwidth, or a local-scale, such as the tunneling rate, is essentially determined by the position and the profile of the initial wave packet. 
\begin{figure}[ht]
\centerline{
    \mbox{\includegraphics[width=0.5\textwidth]{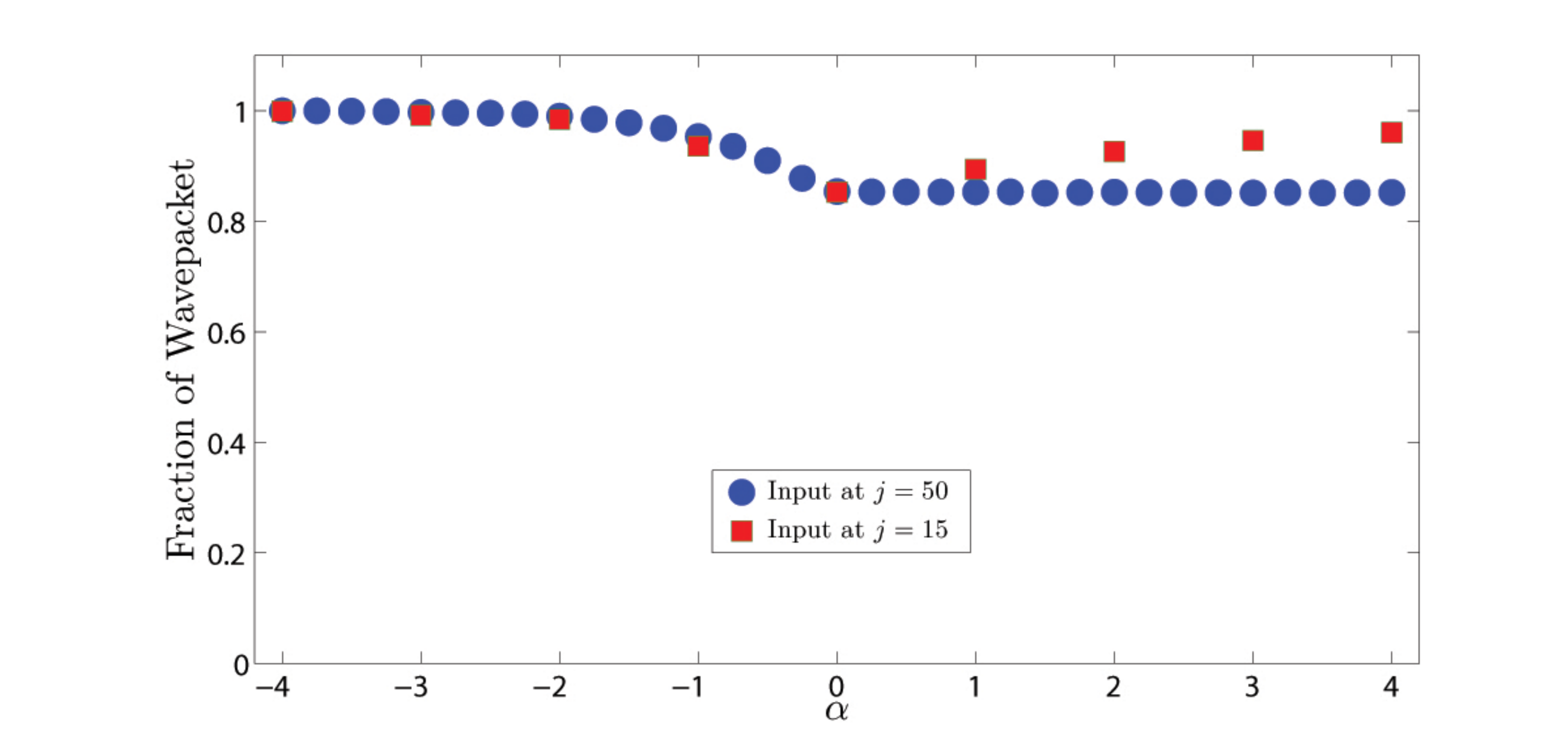}}
    \mbox{\includegraphics[width=0.5\textwidth]{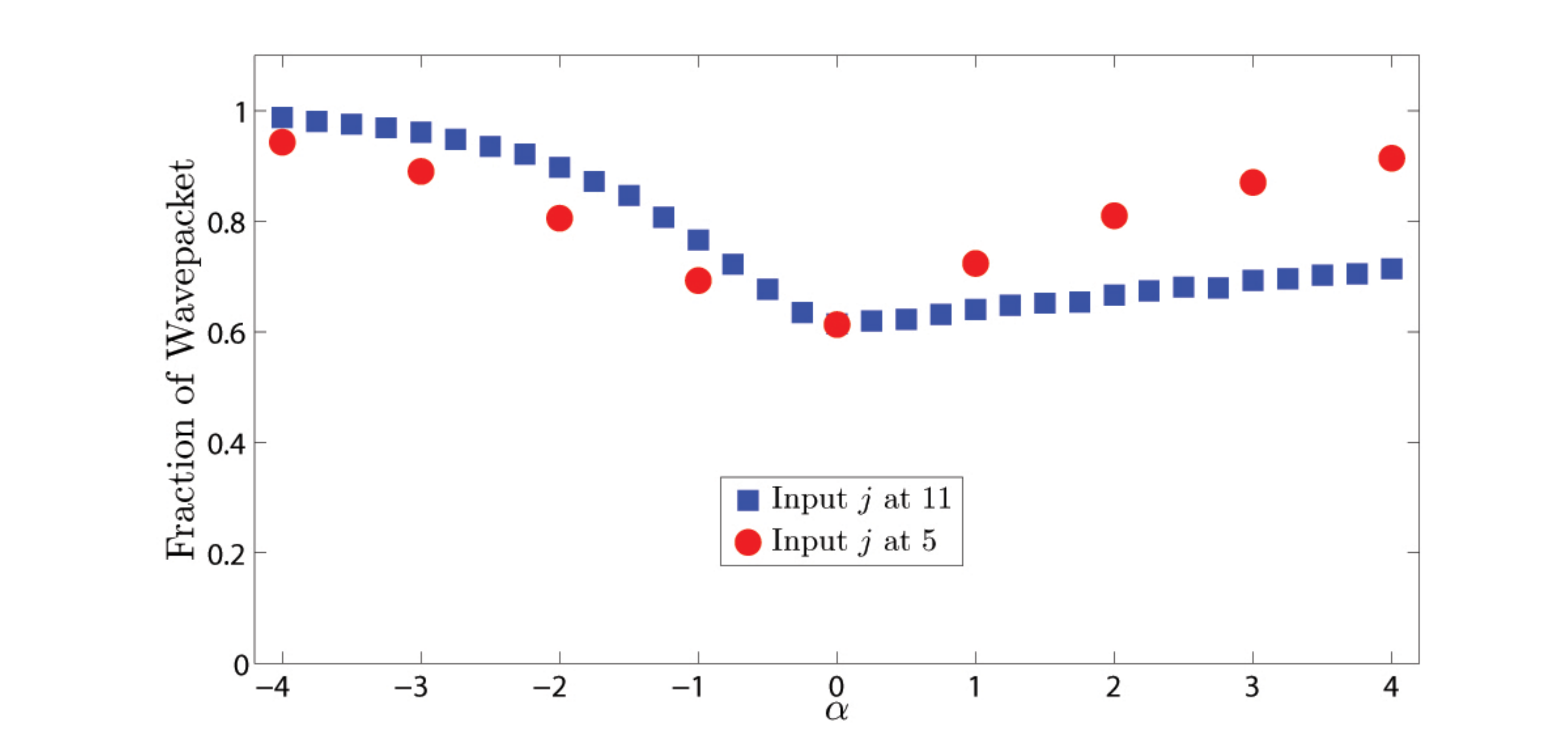}}
  }
\caption{(Color Online) The left panel shows the steady-state intensity localized to the initial waveguide as a function of $\alpha$ for an array with $N=100$ waveguides, $N_r=10^6$ disorder realizations, and input state $|\psi(0)\rangle=|m_0\rangle$. The input state is given by $m_0=50$ (blue circles) and $m_0=15$ (red squares), and the disorder strength is higher than the bandwidth $\hbar\sigma/\Delta_\alpha^{(0)}=3$. We see that the localized fraction is weakly dependent upon $\alpha$. The right panel shows corresponding results for an array with $N=37$ waveguides, $N_r=10^5$ disorder realizations, and a weaker disorder $\hbar\sigma/\Delta_\alpha^{(0)}=1$. We see that the localized fraction as a function of $\alpha$ depends acutely on different initial input states, $m_0=11$ (blue squares) and $m_0=5$ (red squares), when the input locations are relatively close to the boundary.} 
\label{fig:loc}
\end{figure}

The left panel in Fig.~\ref{fig:loc} shows the fraction of total intensity that remains in the initial waveguide as a function of $\alpha$ for $N=100$, initial waveguides $m_0=50$ (blue circles) and $m_0=15$ (red squares), and disorder strength $\hbar\sigma/\Delta^{(0)}_\alpha=3$. When the initial waveguide is near the center of the array, for $\alpha<0,$ we see that the localized fraction rapidly saturates as $|\alpha|$ increases, whereas for $\alpha \geq 0,$ the localized fraction is approximately independent of $\alpha$. On the other hand, when the initial waveguide is near the edge, $m_0=15$, we see a clear dependence of the localized fraction on the tunneling exponent $\alpha$. Note that for a weak disorder, $\hbar\sigma/\Delta^{(0)}_\alpha\ll 1$, the saturation of the localized fraction occurs for $\alpha < 0$, whereas for $\alpha >0$ the disorder has minimal effect and the average intensity at a particular waveguide is nearly uniform, given by $I_a=1/N$. The right panel in Fig.~\ref{fig:loc} shows the localized fraction as a function of $\alpha$ for a smaller disorder  strength $\hbar\sigma/\Delta^{(0)}_\alpha=1$, with $N=37$  and different input waveguide locations $m_0=11$ (blue circles) and $m_0=5$ (red squares). In general, we see that the $\alpha$-dependence of the steady-state localized fraction is sensitive to the proximity of the initial waveguide location to the edge of the array. This result shows that for non-uniform waveguide arrays, the disorder-averaged steady-state intensity profile for a given input state $|m_0\rangle$ is primarily determined by the competition between disorder and the {\it local} tunneling rate $C(m_0)$, instead of disorder and the clean-system bandwidth $\Delta^{(0)}_\alpha$. When $\hbar\sigma\gg\Delta^{(0)}_\alpha$, we find that the localized intensity fraction in the initial waveguide is independent of $\alpha$ since the ballistic component of the intensity time-evolution is completely suppressed by the disorder. Finally, we note that the results displayed in Fig.~\ref{fig:loc} have been systematically checked by us to confirm that the effects shown are not due to the size of the lattice, and are due to the relative proximity of the input wavepacket to the boundary. 

\begin{figure}[h!]
\begin{center}
\includegraphics[width=0.8\textwidth]{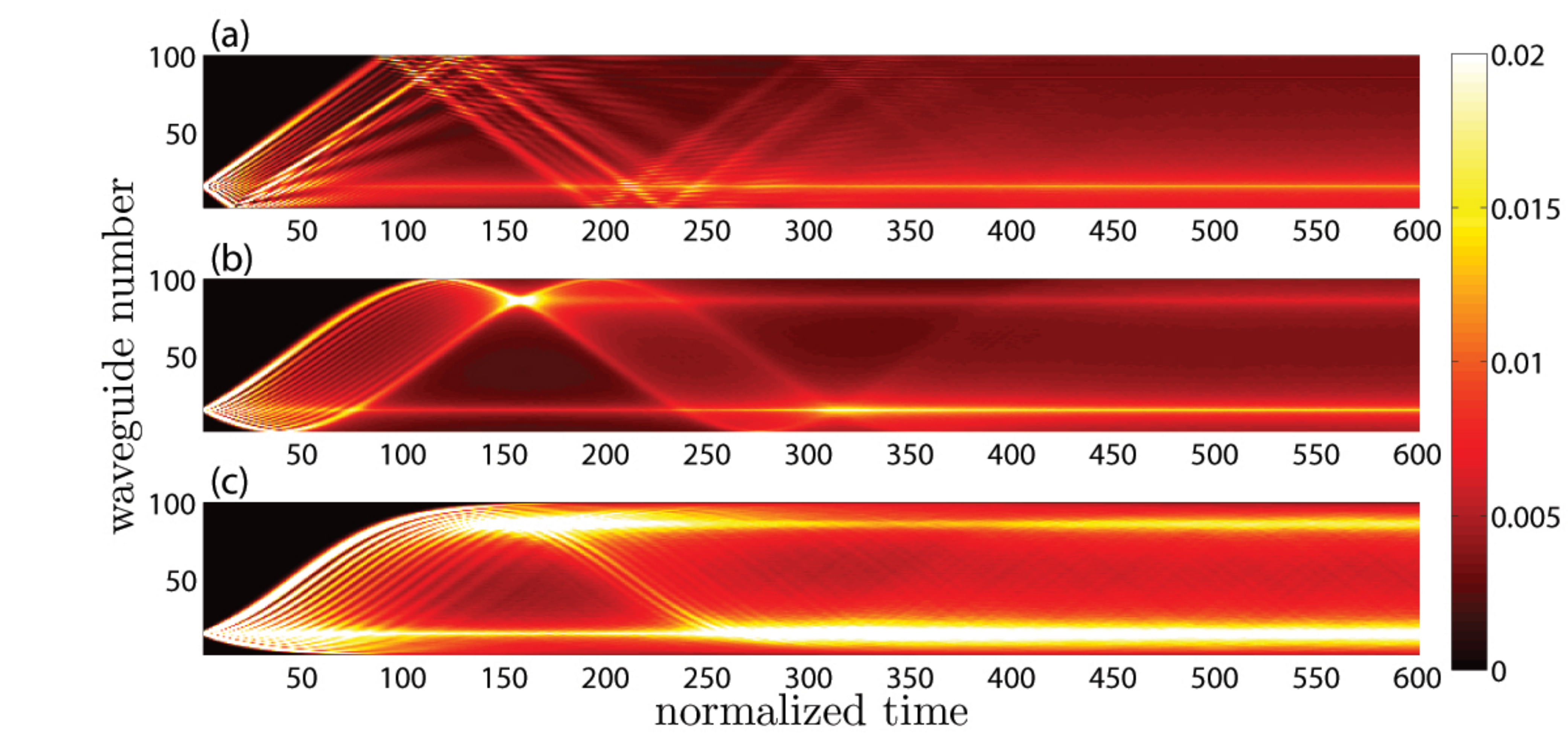}
\caption{(Color Online)  $\alpha$-dependent disorder averaged intensity $I(j,t)$ for an input state $|\psi(0)\rangle=|m_0=15\rangle$ in an array with $N=100$ waveguides, a weak disorder $\hbar\sigma/\Delta^{(0)}_\alpha=0.05$, and $N_r=10^6$ disorder realizations. The horizontal axis in each panel indicates time in units of $\tau_\alpha$. Panel (a) shows exponential localization with a single peak at $m_0$ for an array with uniform tunneling $\alpha=0$. Panels (b) and (c) show corresponding results for $\alpha=1$ and $\alpha=2$ respectively. In each case, a ballistic expansion and reconstruction is followed by emergence of steady-state intensity profile $I(j)$ that has two peaks, one at the input waveguide $m_0$ and the other at its mirror-symmetric counterpart, $N+1-m_0$. The relative weights at the two peaks can be tuned by the varying the weak disorder and the input position $m_0$.}
\label{fig-alphaloc}
\end{center}
\end{figure}
We now consider $\alpha$-dependence of the disorder-averaged intensity profile $I(j,t)$ for a weak disorder $\hbar\sigma/\Delta^{(0)}_\alpha=0.05$ when the input state is localized in waveguide $m_0=15$ near the edge of an $N=100$ waveguide array, $|\psi(0)\rangle=|m_0\rangle$. Figure \ref{fig-alphaloc} shows that initially, the wave packet ballistically spreads and then, at times $t/\tau_\alpha\geq 300\gg 1$, develops a steady-state intensity profile that reflects its localization. The top panel (a) shows that for a uniform array, $\alpha=0$, the intensity is maximum in the initial waveguide, and decays monotonically away from it; note that the maximum intensity $I(m_0)\sim 0.02$ is approximately twice the average intensity $I_a=1/N=0.01$. The middle panel (b) shows that for $\alpha=1$, the disorder-averaged intensity profile changes from periodic reconstruction to localization as time increases. In contrast with the uniform tunneling $\alpha=0$ case, the steady-state intensity $I(j)$ in this case shows two peaks, with different weights, at mirror-symmetric positions $m_0=15$ and $N+1-m_0=86$. The bottom panel (c) shows that for $\alpha=2$, the intensity profile has two sharply defined peaks at mirror symmetric locations. We emphasize that the existence of two peaks in the disorder-averaged intensity is a generic feature of parity-symmetric tunneling function $C_\alpha$, and the ratio of weights at the two peaks can be varied by changing the disorder strength. Thus, in contrast with the exponential localization in a uniform waveguide array, the localization profile of a wave packet in an array with $\alpha\neq 0$ can be varied dramatically by appropriate choice of the input waveguide $m_0$ and disorder strength $\hbar\sigma$. When the disorder strength is increased further, $\hbar\sigma/\Delta^{(0)}_\alpha\geq 1$, the twin-peak structure disappears and the localization profile approaches an exponential as is expected~\cite{Thompson2010}.

\begin{figure}[ht]
\centerline{
    \mbox{\includegraphics[width=0.5\textwidth]{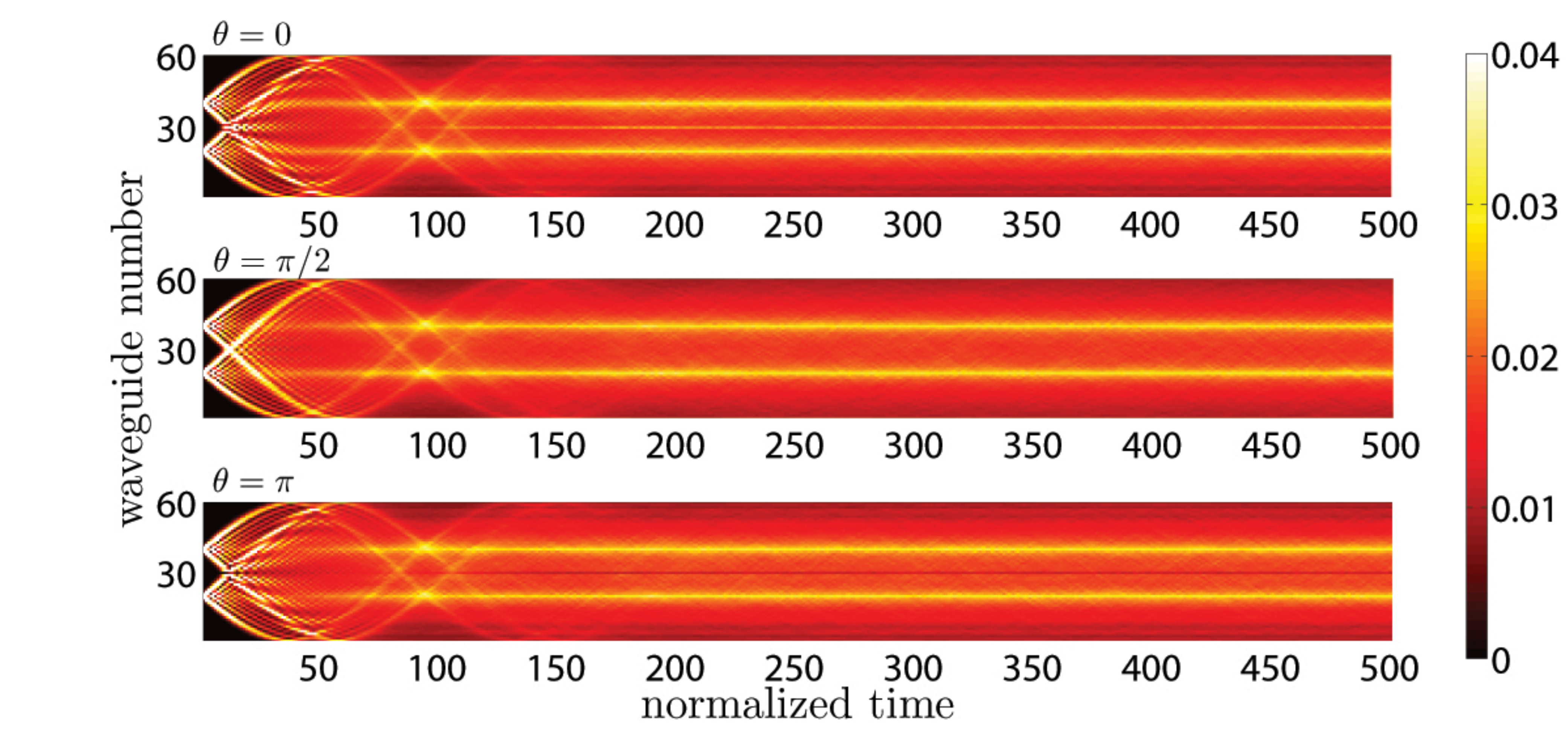}}
    \mbox{\includegraphics[width=0.5\textwidth]{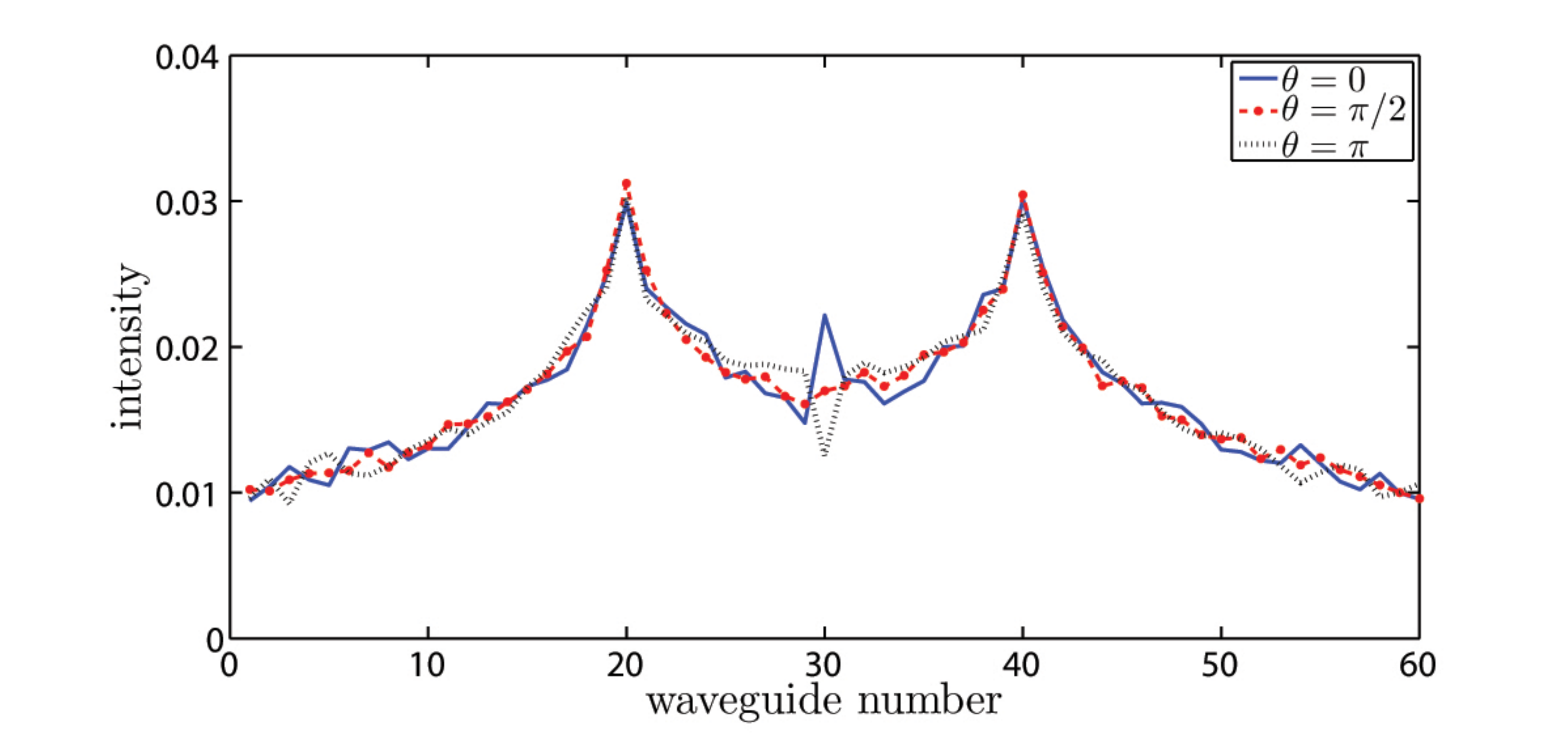}}
  }

\caption{(Color Online) Left-hand panel shows intensity $I(j,t)$ as a function of $\theta$ for an array with $N=60$ waveguides, disorder $\hbar\sigma/\Delta_\alpha^{(0)}=0.05$, and $N_r=10^5$ disorder realizations. The initial input state is $|\psi(0)\rangle=(|20\rangle+e^{i\theta}|40\rangle)/\sqrt{2}$. The top, middle, and bottom panels correspond to $\theta=0$, $\theta=\pi/2$ and $\theta=\pi$ respectively. The interference structure at short times $t/\tau_\alpha\lesssim 100$ is replaced by a steady-state double-peak intensity profile at times $t/\tau_\alpha\geq 300$. The peak intensity is twice the average intensity $I_a=1/N=0.0167$ per waveguide. Right-hand panel shows the corresponding steady-state intensity $I(j)$ at $t/\tau_\alpha=600$ as a function of phase $\theta$. The intensity near the waveguide array center shows enhancement for $\theta=0$ (blue solid line)  and suppression for $\theta=\pi$ (black dotted line) when compared with the corresponding intensity for $\theta=\pi/2$ (red dashed line). As expected, this intensity difference vanishes for a strong disorder.}
\label{fig:phaseloc}
\end{figure}

The results so far have been on the evolution of a single particle that is input to a single waveguide.  In recent years, motivated primarily by the desire to understand the quantum correlations between indistinguishable particles, there have been studies on the evolution of two particles in waveguide lattices.  Bromberg and co-workers have investigated quantum and classical correlations when two photons are coupled to either the same waveguide or to adjacent waveguides~\cite{Bromberg2010}, and Lahini and co-workers~\cite{Lahini2010} have studied a similar problem in disordered lattices.  More recently, coupling of two phase-displaced wavepackets into adjacent waveguides has been used to propose a form of the quantum ratchet~\cite{Thompson2011}.  Motivated by these works, we next explore the effects of disorder and $\alpha$-dependent clean-system spectrum on an input wave packet that is localized to two waveguides with a relative phase $\theta$ between them, $|\psi_\theta(0)\rangle=\left(|p\rangle + e^{i\theta} |q\rangle\right)/\sqrt{2}$, with $1\leq p, q\leq N$. The phase $\theta$ determines the time-evolved intensity $I(j,t)$ in a clean system and for $\alpha=1$, the phase information can only be extracted from intensity measurements in certain time-windows~\cite{Joglekar2011}. Figure~\ref{fig:phaseloc} shows the interplay between the phase $\theta$ and weak disorder, and their effect on the disorder-averaged steady state intensity profile for a non-uniform waveguide array with $N=60$, $\alpha=1$, and an initial state with $p=20$ and $q=N-p=40$. 

The left-hand panel shows the disorder-averaged intensity $I(j,t)$ for $\theta=0$ (top panel), $\theta=\pi/2$ (middle panel), and $\theta=\pi$ (bottom panel) with a weak disorder, $\hbar\sigma/\Delta^{(0)}_\alpha=0.05$, and $N_r=10^5$. At short times $t/\tau_\alpha\lesssim 300$, the intensity shows clear signatures of $\theta$-dependent interference and reconstruction due to equidistant energy levels of a clean $\alpha=1$ system. At large times $t/\tau_\alpha\geq 300$, a steady-state intensity profile with two equal-weight, broad peaks near waveguides $p=20$ and $q=N-p=40$ emerges. A visual inspection of the three intensity profiles shows that the phase information is encoded in the steady-state intensity near the center of the waveguide array. The right-hand panel shows the steady-state intensity profile $I(j)$ at time $t/\tau_\alpha=600$ as a function of the phase $\theta$; recall that the average intensity is given by $I_a=1/N=0.0167$. We see clearly that $\theta=0$ (blue solid line) and $\theta=\pi$ (black dotted line) are marked by increased and suppressed intensity at the center of the waveguide array respectively, compared to the intensity value for $\theta=\pi/2$ (red dashed line). Thus, the phase information, accessible only in certain time windows in clean system, can be extracted from the disorder-averaged steady-state intensity profile; as the disorder gets stronger, this phase information disappears. 

These results show that a complete characterization of the disorder-induced steady-state intensity profile and the time at which steady-state is achieved is a nontrivial problem. In particular, we find that the steady-state intensity $I(j)$ depends acutely on the size and relative phase $\theta$ of the initial input state $|\psi_\theta(0)\rangle$, the tunneling function $\alpha$, and the proximity of the input state with the boundaries.  


\section{Intensity correlations in the presence of disorder: Hermitian vs. non-Hermitian case}
\label{sec:hbt}

\begin{figure}[ht]
\centering
\includegraphics[width=1\textwidth]{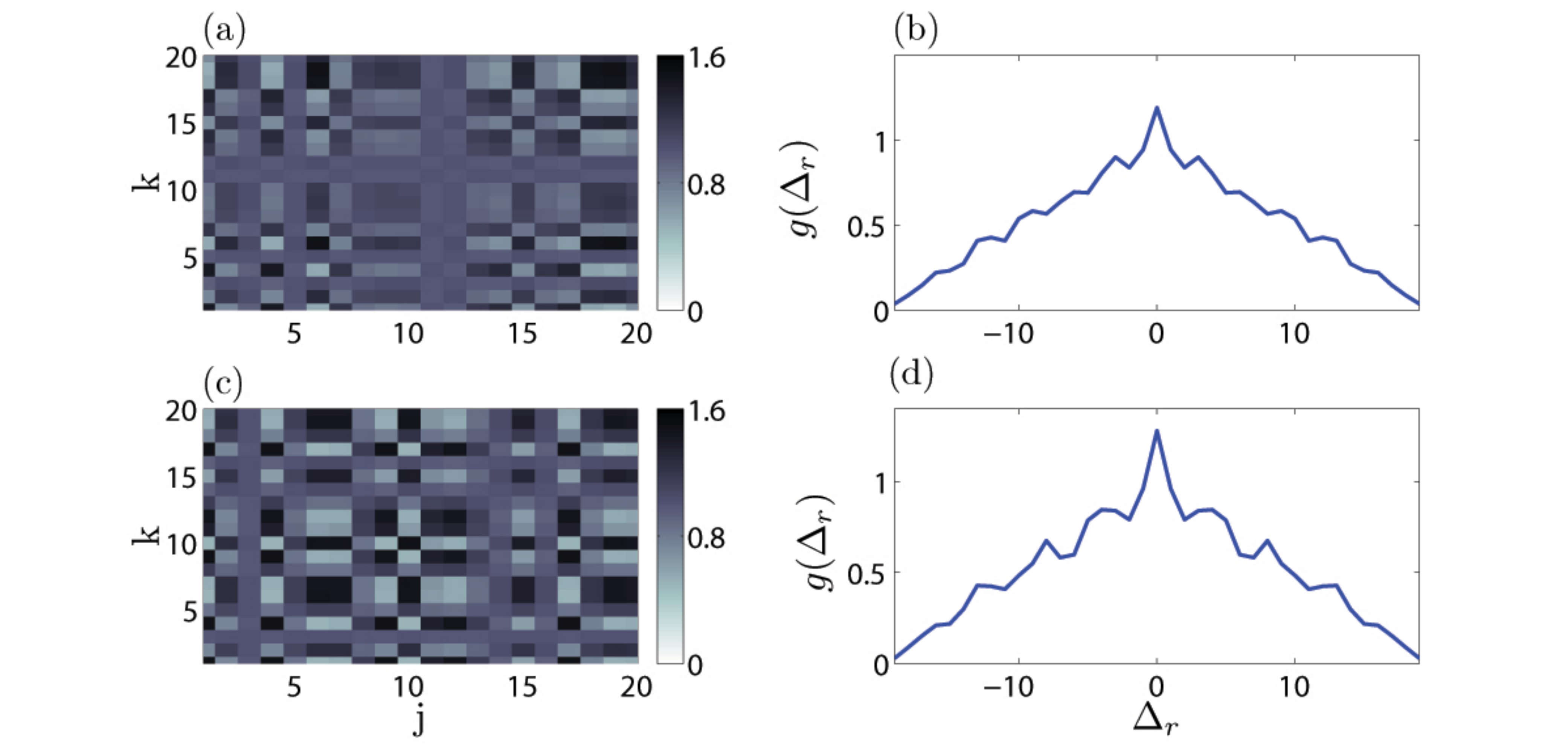}
\caption{(Color Online) Left-hand column shows the disorder-averaged, steady-state, classical correlation matrix $\Gamma_{jk}$ for a uniform array with $N=20$ waveguides and a weak disorder $\hbar \sigma /\Delta_{\alpha}=0.02$; the results are averaged over $N_r=10^4$ disorder realizations. Panel (a) shows the matrix for on-site, $\mathcal{PT}$-symmetric, non-Hermitian disorder; panel (b) shows the matrix for tunneling-rate, Hermitian disorder. The right-hand column shows the correlation functions $g(\Delta_r)$ extracted from the steady-state, classical correlation matrix for on-site, $\mathcal{PT}$-symmetric disorder, panel (c), and off-diagonal, tunneling disorder, panel (d). The similarity between the two results shows that the particle-hole symmetry of the disordered energy spectrum is instrumental to the correlation function properties.}
\label{fig:hbt}
\end{figure}

In the previous section, we focused on localization due to a Gaussian, Hermitian disorder in the on-site potential because the localization intensity profile is independent of the disorder origin (on-site potential or tunneling rates) and disorder probability distribution as long as different distributions have zero mean and the same variance~\cite{Thompson2010,Lahini2011}. In contrast to this, higher-order intensity correlations provide a clue into the origin of the disorder; this is because an on-site disorder destroys the particle-hole symmetry of the clean-system energy spectrum whereas a tunneling-rate disorder preserves it~\cite{Lahini2011}. In this section, we explore the effects of a non-Hermitian, $\mathcal{PT}$-symmetric, on-site disorder on the intensity-intensity correlations. To this end, we modify the Hamiltonian $\hat{H}_\alpha$ to include balanced gain ($i|\gamma|$) and loss ($-i|\gamma|$) terms~\cite{Makris2008,Kottos2010,Ruter2010a},
\be
\label{eq:pt}
\hat{H}_\mathcal{PT}=\hat{H}+\sum_{m=1}^{N/2} i\gamma_m ( a^{\dagger}_m a_m-a^{\dagger}_{\bar{m}} a_{\bar{m}})
\ee
where $1\leq m\leq N/2$ is the position of gain waveguide and $\bar{m}=N+1-m$ denotes the index of the loss waveguide. Although $\hat{H}_\mathcal{PT}$ is not Hermitian, it has purely real eigenvalues and relatively strong $\mathcal{ PT}$ symmetric phase for $\alpha>0$~\cite{Scott2011}. We choose random, uniformly distributed on-site impurities $\gamma_m$ for $1\leq m\leq N/2$; such a choice of random, on-site impurities leads to a purely real and  {\it particle-hole symmetric} spectrum~\cite{Joglekar2010}. Motivated by this result, we compare and contrast the intensity-intensity correlations due to $\mathcal{PT}$-symmetric disorder and off-diagonal disorder in tunneling rates. We recall that the position-dependent tunneling profile is given by 
Eq.(\ref{eqn:tunnelingrate}), $C_\alpha(j)=C_0[j(N-j)]^{\alpha/2}$. We introduce the disorder in the tunneling rate via $C_0\rightarrow C_0[1+\delta(j)]$ where the random, position-dependent change $\delta(j)$ is drawn from a uniform distribution with zero mean and variance $\sigma^2$ and confine ourselves to a weak disorder which ensures that the scale-factor $1+\delta(j)$ is always positive. The disorder-averaged, classical, steady-state correlation matrix is defined as~\cite{Lahini2011}
\be
\label{eq:hbt}
\Gamma_{jk}(t)=\frac{\langle I(j,t) I(k,t)\rangle}{\langle I(j,t)\rangle\langle I(k,t)\rangle}
\ee
where $I(j,t)$ is the intensity profile which depends upon the initial state $|\psi_\theta(0)\rangle$, and $\langle\cdots\rangle$ indicates averaging over different disorder realizations and different relative phases $\theta$ between the light input into the two waveguides $|p\rangle$ and $|q\rangle$. Note that since the $\mathcal{PT}$-symmetric Hamiltonian, Eq.(\ref{eq:pt}), is not Hermitian, the corresponding time-evolution operator is not unitary, and therefore the total intensity $\sum_{j} I(j,t)$ may not be conserved. For the purpose of illustration, we show the results for a uniform ($\alpha=0$) $N=20$ waveguide array with initial state $|\psi_\theta(0)\rangle=(|N/2-1\rangle+|N/2\rangle)/\sqrt{2}$. The disorder strength is $\hbar \sigma /\Delta_{\alpha}=0.02$ and $N_r=10^4$. The left-hand column in Fig.~\ref{fig:hbt} show the disorder-averaged, steady-state matrix $\Gamma_{jk}$ for $\mathcal{PT}$-symmetric, on-site disorder, panel (a), and tunneling disorder, panel (c). 

Traditionally, however, instead of the entire correlation matrix $\Gamma_{jk}$ with $N^2/2$ independent entries, one considers the correlation function $g(\Delta_r)=N^{-1}\sum_{j=1}^N\Gamma_{j,j+\Delta_r}=g(-\Delta_r)$ with $N$ independent entries. This correlation function is able to distinguish between on-site potential disorder and off-diagonal tunneling-rate disorder~\cite{Lahini2011}. The right-hand column in Fig.~\ref{fig:hbt} shows correlation functions extracted from the steady-state matrix $\Gamma_{jk}$ for on-site $\mathcal{PT}$-symmetric disorder, panel (b), and off-diagonal tunneling rate disorder, panel (d). The similarity between the two correlation functions, and their stark contrast with the corresponding correlation function for a Hermitian on-site disorder~\cite{Lahini2011}, shows that the particle-hole symmetry in the spectrum of a disordered system, rather than the origin of the disorder, is instrumental in determining the correlation function properties. 

\section{Discussion}
\label{sec:disc}
In this paper, we have explored the effects of disorder in waveguide arrays with non-uniform tunneling and non-Hermitian, on-site, $\mathcal{PT}$-symmetric disorder, by focusing on the behavior of disorder-averaged, steady-state intensity profile and intensity-intensity correlations. 

Broadly, we found that the intensity profile $I(j,t)$ is acutely sensitive to the tunneling function $C_\alpha$, the initial input state $|\psi_\theta(0)\rangle$, its width and its proximity to one of the edges, and does not necessarily result in an exponential localization profile that is well known in uniform waveguide arrays~\cite{Lahini2008}. In particular, we found that when $\alpha\geq 1$ and input state is localized to waveguide $m_0$, a weak disorder suppresses the wave packet reconstruction and causes localization of wave packet in its mirror-symmetric waveguide $N+1-m_0$, as well as its initial waveguide $m_0$. We also found that the steady-state intensity $I(j)$ encodes the phase-information of the initial state $|\psi_\theta(0)\rangle$ for a weak disorder. In all cases, a strong disorder $\hbar\sigma/\Delta_\alpha^{(0)}\gg 1$, results in a localized intensity profile that is virtually identical to the initial intensity profile. 

Although the localization intensity profile is insensitive to the origin of the disorder, on-site or off-diagonal, higher-order intensity correlations depend upon it; in particular, Hermitian on-site and off-diagonal disorders lead to qualitatively different intensity correlation function $g(\Delta_r)$~\cite{Lahini2011}. Here, we have shown that a Hermitian disorder in the tunneling rate, and a non-Hermitian, $\mathcal{PT}$-symmetric, on-site disorder result in nearly identical correlation functions. Thus, the behavior of the correlation function $g(\Delta_r)$ can be traced to the presence or absence of particle-hole symmetry in the spectrum of a disordered Hamiltonian. 

In this paper, we have ignored the effect of Kerr nonlinearity that becomes relevant at high electric field amplitudes, and that affects the disorder-induced localization of positive and negative energy states differently~\cite{Lahini2008}. Even in the absence of this nonlinearity, our results show that the properties of localization in tunable waveguide arrays, including the dependence 
of saturation time $T_s$ beyond which a steady-state intensity profile emerges on the array parameters, are barely explored. Finally, we find that the localization of light in a finite lattice can have a complicated dependence on $\alpha$, the array size and the disorder, and further work is necessary to identify the interplay among these factors in determining the behavior of light in a coupled array of waveguides.

\begin{acknowledgments}
C.T. was supported by a GAANN fellowship from the US Department of Education awarded to G.V.
\end{acknowledgments}


%

\end{document}